\documentclass[aps,preprint,prd,nofootinbib]{revtex4}
\usepackage{subfigure}
\usepackage{array}
\usepackage{graphicx}
\usepackage{caption}
\usepackage{subcaption}
\usepackage{amsmath}
\usepackage{amssymb}
\usepackage{mathrsfs}
\usepackage{slashed}

\begin{document}

\title{The electromagnetic Dalitz decay of vector double heavy mesons}
\author{Zhi-Hui Wang$^{[1],[2]}$\footnote{zhwang@nmu.edu.cn}, Tianhong Wang$^{[3]}$\footnote{thwang@hit.edu.cn}, Yue Jiang$^{[3]}$\footnote{jiangure@hit.edu.cn}, Guo-Li Wang$^{[4],[5]}$\footnote{wgl@hbu.edu.cn}\\}
\address{$^1$ Key Laboratory of Physics and Photoelectric Information Functional Materials, North Minzu University, Yinchuan, 750021, China,\\
$^2$School of Electrical and Information Engineering, North Minzu University, Yinchuan, 750021, China,\\
$^3$School of Physics, Harbin Institute of
Technology, Harbin, 150001, China,\\
$^4$Department of Physics, Hebei University, Baoding, 071002, China,\\
$^5$Hebei Research Center of the Basic Discipline for Computational Physics, Baoding 071002, China}

 \baselineskip=20pt

\begin{abstract}
In this paper, we study the electromagnetic Dalitz decay processes
$V \to P l^-l^+$ ($l = e,\mu$), where $V$ and $P$ are vector and
pseudoscalar double-heavy mesons, respectively.
Within the improved
Bethe-Salpeter method, we calculate the transition form factors, the
dilepton invariant-mass spectra, the partial decay widths, and the branching
fractions for the charmonium, bottomonium, and $B_c$ systems.
Our results
$\mathcal{B}(J/\psi\to\eta_c e^-e^+) = 1.15\times10^{-4}$ and
$\mathcal{B}(\psi(2S)\to\eta_c e^-e^+) = 2.62\times10^{-5}$ are consistent with
the latest BESIII measurements.
The partial widths of the $B_c$ channels are predicted to be of
order $10^{-10}$~GeV.
These predictions will provide useful information for future experimental searches for these decay channels at
BESIII, Belle~II, and LHCb.
 \vspace*{0.5cm}


\end{abstract}

\maketitle
\section{Introduction}
The electromagnetic(EM) Dalitz decay processes $h_1\to h_2l^-l^+$,
where $h_1$ and $h_2$ are double heavy mesons can help us
to understand the structure of the heavy mesons and the fundamental mechanisms of the interactions between photons
and the heavy mesons.
LHCb Collaboration observed the decays $\chi_{c1,c2}\to J/\psi \mu^+\mu^-$ in 2017~\cite{lhcb1}.
Then, BESIII Collaboration observed the decays $\psi(3686)\to \chi_{cJ} e^+e^- $ and $\chi_{cJ}\to J/\psi e^+e^-$ for the first time in 2017~\cite{BESIII3686}.
BESIII Collaboration reported the decays $\chi_{cJ}\to J/\psi \mu^+\mu^-$ through the radiative decays
$\psi(3686)\to \gamma\chi_{cJ}$ in 2019,
the corresponding branching fractions were $\mathcal B(\chi_{c1}\to J/\psi \mu^+\mu^-)=(2.51\pm0.18\pm0.20)\times 10^{-4}$
and $\mathcal B(\chi_{c2}\to J/\psi \mu^+\mu^-)=(2.33\pm0.18\pm0.29)\times 10^{-4}$.
But the decay $\chi_{c0}\to J/\psi \mu^+\mu^-$ was not observed~\cite{BESchiJ}.
The branching fractions were $\mathcal B(\psi(3686)\to \chi_{cJ} e^+e^- )=(11.7\pm2.5\pm1.0)\times 10^{-4},
\ (8.6\pm0.3\pm0.6)\times 10^{-4},\ (6.9\pm0.5\pm0.6)\times 10^{-4}$ with $J=0,1,2$
and $\mathcal B(\chi_{cJ}\to J/\psi e^+e^-)=(1.51\pm0.30\pm0.13)\times 10^{-4},\ (3.73\pm0.09\pm0.25)\times 10^{-3}
\ (2.48\pm0.08\pm0.16)\times 10^{-3}$ with $J=0,1,2$, respectively.
BESIII Collaboration observed the EM Dalitz decay $\psi(3686)\to \eta_c e^-e^+$ with a significance of 7.9$\sigma$~\cite{BESIII:2022jde},
and the corresponding branching fraction was
$\mathcal B(\psi(3686)\to \eta_c e^-e^+)=(3.77\pm 0.40\pm 0.18)\times 10^{-5}$.
Recently, BESIII Collaboration reported the electromagnetic Dalitz transition
$\mathcal B(J/\psi\to \eta_c e^-e^+)=(0.92\pm 0.03_{stat}\pm 0.10_{syst}\pm0.09_{input})\times 10^{-4}$\cite{BESIII:2026}.

The EM decay processes of heavy mesons have inspired great interest in
both theoretical and experimental research fields of hadronic physics.
Many theoretical methods
have studied the EM decay of heavy mesons~\cite{fujl,Luch,zhukai,liuchuan,zhongxh,zhaoqiang,gaozb,yangmaozhi,peisy,Ridwan,hejk}.
Such as Ref.~\cite{Luch} presented the numerical values of the branching fractions of
exclusive radiative decays $\chi_{cJ}\to J/\psi\mu^+\mu^-$ and $\chi_{bJ}\to \Upsilon\mu^+\mu^-$.
Ref.~\cite{zhaoqiang} calculated the E1 decays of $\psi(3686)$ and $\psi(3770)\to \gamma \chi_{cJ}$
by the non-relativistic effective field theory.
Ref.~\cite{yangmaozhi} studied the electromagnetic Dalitz decays $\psi(\Upsilon)\to \eta_c(\eta_b)l^+l^-$,
and predicted the corresponding partial width.
Ref.~\cite{Ridwan} studied the M1 radiative transitions of excited heavy quarkonia and $B_c$ mesons by
the light-front quark model.
We also calculated the dilepton electromagnetic decays of $\chi_{cJ}(1P)$ by the improved Bethe–Salpeter method,
and got some valuable results~\cite{dilepton}.
However, there are few studies for the EM Dalitz decay of the excited vector double heavy mesons.
$\psi(2S)$ and $\Upsilon(2S)$ have been observed in the experiment~\cite{pdg},
$B^*_c(2S)$ also was reported by LHCb Collaboration~\cite{bcstar2s}.
The study of vector double heavy mesons, not only helps us to better understand the quark structure of the vector double heavy mesons,
but also provides
the useful information to discover and confirm these decay channels in the future.

In this paper we will study the EM Dalitz decays vector double heavy mesons,
which the final states are pseudoscalar double heavy mesons.
The relativistic correction is very important in the double heavy mesons,
the relativistic correction of vector heavy mesons ($\psi(2S)$, $\Upsilon(2S)$ and $B^*_c(2S)$) are larger than that of
the corresponding 1$S$ states,
therefore, a relativistic model is needed.
The improved Bethe-Salpeter (BS) method is
a relativistic framework that describes the bound state with definite quantum number,
the corresponding relativistic wave functions are obtained by solving the full Salpeter equations.
The properties of some
heavy mesons have been calculated by the improved BS method in previous work,
such as
the dilepton electromagnetic decays of $\chi_{cJ}(1P)$~\cite{dilepton},
the decay constants of heavy vector mesons~\cite{vector},
radiative E1 decays of $X(3872)$~\cite{3872},
the electromagnetic decays of $B_c^\pm(2S)$~\cite{bc2s},
the mass spectrum and wave functions of the $B_c$ system~\cite{bc},
the electromagnetic decays of $X(3823)$ as the $\psi_2(1^3D_2)$ state and its radial excited states~\cite{3823},
$\eta_{c2}(^1D_2)$ and its electromagnetic decays~\cite{etac2}.
The relevant results are consistent with the experimental or other theoretical results.
So we describe the vector double heavy mesons using the BS equation in this paper.

The mass of the vector charmonium states ($J/\psi$, $\psi(2S)$)
and bottomonium states ($\Upsilon$, $\Upsilon(2S)$) are below $D\bar D$ and $B\bar B$, respectively,
they have no OZI-allowed strong decays,
the $B^*_c$ and $B^*_c(2S)$ also have no strong decays,
so the EM decays of these vector double heavy mesons are very important.
Thus, we study the EM Dalitz decay processes $V\to Pl^-l^+$ of these vector double heavy mesons in this paper.

The paper is organized as follows.
We give the formulation of the EM Dalitz decay processes $V\to Pl^-l^+$ in Section II;
In Sec. III, we show the numerical results and discussions;
We give the corresponding summary in Sec. IV. Finally, we present the instantaneous BS
equation in the Appendix.

\section{Theoretical calculations}

\begin{figure}[htbp]
\centering
\includegraphics[height=5.5cm]{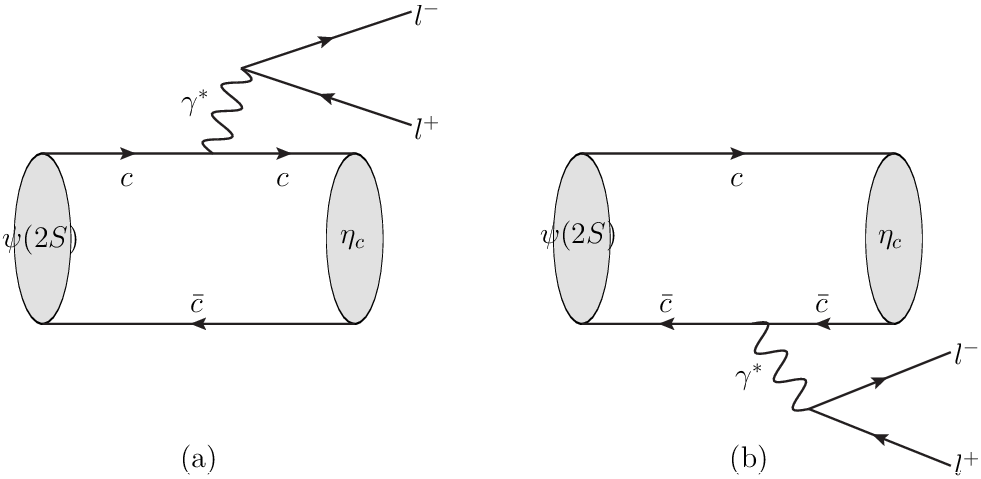}
\caption{\label{feyn} Feynman diagram of $\psi(2S)\to \eta_c l^-l^+$.}
\end{figure}

We study the decay process $V(P)\to P(P_f)+l^-(k_1)+l^+(k_2)$, where $V$ and $P$ are the initial and final mesons, which have the masses $M$ and $M_f$, respectively.
In this section, we will give the formula of the EM Dalitz decay matrix element, which is general for all the decay channels in this paper.
Taking $\psi(2S)\to \eta_c l^-l^+$ as an example,
the Feynman diagrams are shown in Figure.~\ref{feyn}.

The corresponding Feynman amplitude to the diagrams can be written as
\begin{equation}
	\mathcal{M}=\frac{-ie}{Q^2}H^\mu\bar u(k_1)\gamma_\mu v(k_2),
\end{equation}
where $Q^2=(k_1+k_2)^2$ is the invariant mass of the final leptons with $k_1$ and $k_2$ being the momenta of the final negative-charged and positive-charged lepton, respectively;
the leptonic part $\bar u(k_1)\gamma_\mu v(k_2)$ is model independent,which is from the virtual photon.
$H^\mu$ is the hadronic transition matrix element,
according to the Mandelstam formalism and improved BS method,
the hadronic transition matrix element can be formulated as the overlap integral between the wave functions of initial and final mesons,
\begin{equation}
\label{Hmu1}
\begin{aligned}
H^\mu&=\int\frac{d^3q_\perp}{(2\pi)^3}\,{\rm Tr}\Big\{
\bar{\varphi}^{++}_{P_f}(q_\perp+\alpha_2P_{f\perp})(-iQ_1e\gamma^\mu)\varphi_P^{++}(q_\perp)\frac{\slashed{P}}{M} \\
&+\bar{\varphi}^{++}_{P_f}(q_\perp-\alpha_1P_{f\perp})\frac{\slashed{P}}{M}\varphi_P^{++}(q_\perp)(-iQ_2e\gamma^\mu)\Big\},
\end{aligned}
\end{equation}
where $P$ and $P_f$ are the momentum of the initial and final mesons, respectively, $M$ is the mass of the initial meson.
$Q_1$ and $Q_2$ are the electric charges of the quark and antiquark, respectively.
 $\varphi_P^{++}$ and $\varphi_{P_f}^{++}$ are the positive energy parts of the wave functions of the initial and final mesons, respectively. We have used the definition $\overline{\varphi_{P_f}^{++}}=\gamma^0(\varphi_{P_f}^{++})^\dagger\gamma^0$.
 $q$ is the relative momentum of the quark and aniquark in the initial meson with $q_\perp=q-\frac{P\cdot q}{M^2}P$.
 $\alpha_i~(i=1,~2)$ are defined as $\alpha_i=\frac{m_i}{m_1+m_2}$ with $m_1$ and $m_2$ being the masses of quark and antiquark, respectively.

The positive energy part of the wave function for the initial $1^-$ meson is
\begin{equation}
\begin{aligned}
		\displaystyle\varphi^{++}_{1^-}(q_\perp)&=(q_\perp\cdot\epsilon)\left[A_1(q_\perp)+\frac{\not\!P}{M}A_2(q_\perp)
		+\frac{\not\!q_\perp}{M}A_3(q_\perp)+\frac{{\not\!P}{\not\!q_\perp}}{M^2}A_4(q_\perp)\right]\\
		&+ M{\not\!{\epsilon}}\left[A_5(q_\perp)+\frac{\not\!P}{M}A_6(q_\perp)
		+\frac{\not\!q_\perp}{M}A_7(q_\perp)+\frac{{\not\!P}{\not\!q_\perp}}{M^2}A_8(q_\perp)\right],
\end{aligned}
\end{equation}
$\epsilon$ is the polarization vector and $A_i$ are expressed as

\begin{equation}
\begin{aligned}
A_1 &=\frac{1}{2M(m_1\omega_2 + m_2\omega_1)}\left[(\omega_1+\omega_2) q_\perp^2f_3 + (m_1 + m_2) q_\perp^2f_4 + 2M^2\omega_2f_5 - 2M^2m_2f_6\right],\\
A_2 &= \frac{1}{2M(m_1\omega_2 + m_2\omega_1)}\left[(m_1-m_2) q_\perp^2f_3 + (\omega_1 - \omega_2)q_\perp^2f_4 - 2M^2m_2f_5 + 2M^2\omega_2f_6\right],\\
A_3 &= \frac{1}{2}\left[f_3 + \frac{m_1 + m_2}{\omega_1+\omega_2}f_4 -\frac{2M^2}{m_1\omega_2 + m_2\omega_1}f_6\right],\\
A_4 &= \frac{1}{2}\left[\frac{\omega_1+\omega_2}{m_1+m_2}f_3 + f_4 -\frac{2M^2}{m_1\omega_2+m_2\omega_1}f_5\right],\\
A_5 &= \frac{1}{2}\left[f_5 -\frac{\omega_1+\omega_2}{m_1+m_2}f_6\right], \qquad
A_6 = \frac{1}{2}\left[-\frac{m_1+m_2}{\omega_1+\omega_2}f_5 + f_6\right],\\
A_7 &= \frac{M}{2}\frac{\omega_1-\omega_2}{m_1\omega_2+m_2\omega_1}\left[f_5 - \frac{\omega_1+\omega_2}{m_1+m_2}f_6\right],\\
A_8 &= \frac{M}{2}\frac{m_1+m_2}{m_1\omega_2 + m_2\omega_1}\left[-f_5 +\frac{\omega_1+\omega_2}{m_1+m_2}f_6\right],
\end{aligned}
\end{equation}
where $\omega_i=\sqrt{m_i^2+q^2}$ with $i=1,2$, $f_is$ are the wave functions of the initial vector mesons which can be obtained by solving the BS equations.

For the final $0^-$ meson, the positive energy part of the wave function is
\begin{eqnarray}
	\varphi^{++}_{0^-}(q_{f\perp})=\left[A'_1(q_{f\perp})+\frac{{\not\!P_f}}{M_f}A'_2(q_{f\perp})
		+\frac{\not\!{q}_{f\perp}}{M_f}A'_3(q_{f\perp})+\frac{\not\!{P_f}\not\!{q}_{f\perp}}{M_f^2}A'_4(q_{f\perp})\right]\gamma_5,
\end{eqnarray}
where $M_f$ is the mass of the final meson, $q_{f\perp}=q_\perp\pm\alpha_iP_{f\perp}$,
$A'_i$ are expressed as
\begin{equation}
\begin{aligned}
	&A'_1= \frac{M_f}{2}\left[\frac{\omega'_1+\omega'_2}{m'_1+m'_2}f'_1+f'_2\right],
	~~~A'_2=\frac{M_f}{2}\left[f'_1+\frac{m'_1+m'_2}{\omega'_1+\omega'_2}f'_2\right],\\
	&A'_3=-\frac{M_f(\omega'_1-\omega'_2)}{m'_1\omega'_2+m'_2\omega'_1}A'_1,
		 ~~~~~~~~~A'_4=-\frac{M_f(m'_1+m'_2)}{m'_1\omega'_2+m'_2\omega'_1}A'_1,
\end{aligned}
\end{equation}
where $f'_is$ are the wave functions of the final pseudoscalar mesons,
$m'_1$ and $m'_2$ are the masses of quark and antiquark for the final meson,
$\omega'_i=\sqrt{m^{\prime2}_i+q_f^2}$ with $i=1,2$.

After finishing the integral, the hadronic transition matrix element $H^\mu$ is written as
\begin{equation}\label{Hmu2}
	H^\mu=\epsilon^{\mu\nu\alpha\beta}\epsilon_\nu P_\alpha P_{f\beta}t,
\end{equation}
where $t$ is the form factor. The corresponding hadronic transition matrix element is
\begin{equation}\label{M}
	\mathcal{M}=\frac{-ie}{Q^2}t\epsilon^{\mu\nu\alpha\beta}\epsilon_\nu P_\alpha P_{f\beta}\bar u(k_1)\gamma_\mu v(k_2).
\end{equation}
The decay width can be written as
\begin{equation}\label{Gamma}
	\Gamma=\frac{1}{2M}\int\frac{{\rm d}^3P_f}{(2\pi)^32E_f}\frac{{\rm d}^3P_2}{(2\pi)^32E_2}\frac{{\rm d}^3P_3}{(2\pi)^32E_3}(2\pi)^4\delta^{(4)}(P-P_f-P_2-P_3)\overline{|\mathcal{M}|^2},
\end{equation}
where $E_f$ is the energy of the final mesons, $E_2$ and $E_3$ are the energies of the final charged leptons.
In order to get the invariant mass spectra of final leptons,
the relation of the invariant mass spectra of final leptons and decay width can be expressed as
\begin{equation}\label{Gamma1}
	\frac{{\rm d}\Gamma}{{\rm d}Q}=\frac{1}{256\pi^3M^3}\frac{1}{Q}\sqrt{\lambda(M^2,M_f^2,Q^2)\lambda(Q^2,M_2^2,M_3^2)}\int{\rm d}\cos\theta\bar{|\mathcal{M}|^2},
\end{equation}
where $M_2$ and $M_3$ are the mass of the final leptons, respectively.
$\theta$ is the angle between the final lepton and the final meson in the center of mass system (C. M. S) of the lepton pair.
The form of $\lambda$ can be written as:
\begin{equation}
	\lambda(a,b,c)=a^2+b^2+c^2-2ab-2bc-2ac.
\end{equation}

\section{Numerical results and discussions}

In order to get the wave functions and the mass spectra of the initial and final mesons,
we choose the input parameters as: $m_b=4.96$ GeV, $m_c$=1.62 GeV,
by solving the BS equation, we obtain the mass of mesons:
$M_{J/\psi}=3.096$ GeV, $M_{\psi(2S)}=3.686$ GeV, $M_{\Upsilon(1S)}=9.460$ GeV,
$M_{\Upsilon(2S)}=10.023$ GeV, $M_{\eta_c}=2.984$ GeV, $M_{\eta_b}=9.398$ GeV~\cite{mass1}.
$M_{B_c^\ast(1S)}=6.333$ GeV, $M_{B_c^\ast(2S)}=6.912$ GeV, $M_{B_c}=6.276$ GeV~\cite{bc2s}.
The corresponding width of the mesons are shown as:
$\Gamma_{J/\psi}=9.26\times10^{-5}$~GeV, $\Gamma_{\psi(2S)}=2.86\times10^{-4}$~GeV,
$\Gamma_{\Upsilon(1S)}=5.40\times10^{-5}$~GeV, $\Gamma_{\Upsilon(2S)}=3.20\times10^{-5}$~GeV~\cite{pdg}.

\begin{figure}[htbp]
\centering
\subfigure[]{\includegraphics[scale=0.28]{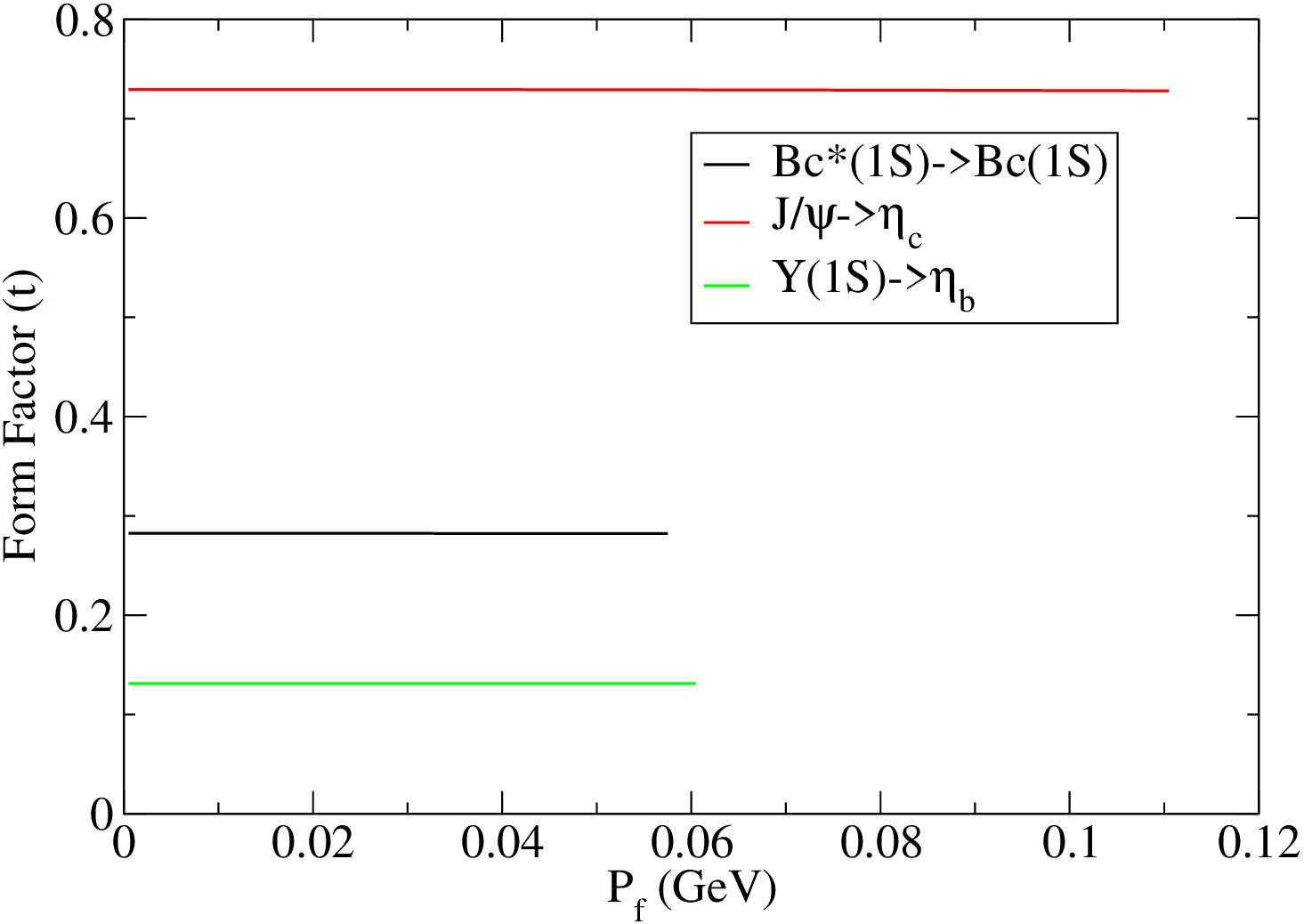}}
\hspace{8mm}
\subfigure[]{\includegraphics[scale=0.28]{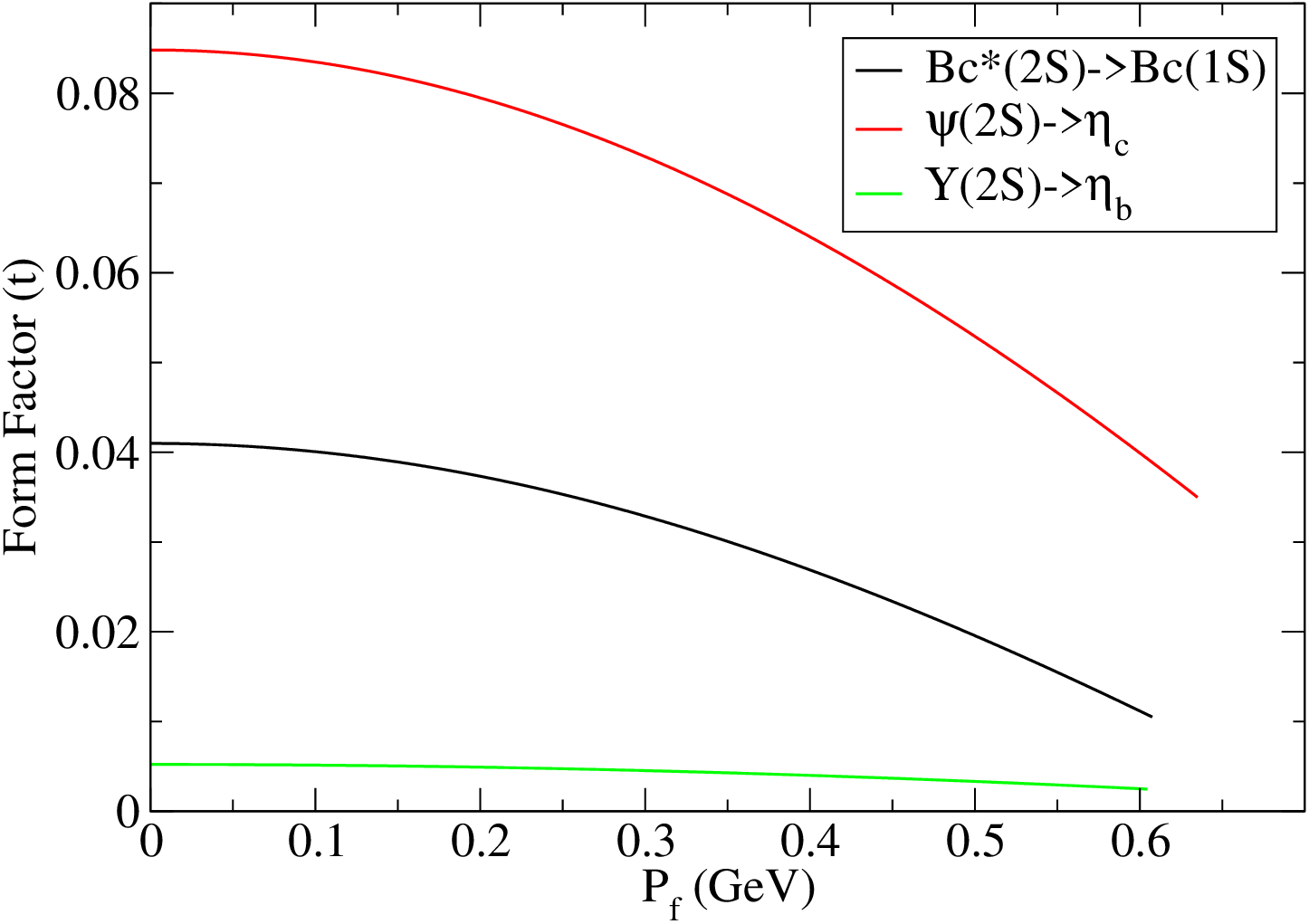}}
\caption{\label{formfactor}Form factors of Dalitz decay of vector heavy mesons.}
\end{figure}

\begin{figure}[htbp]
\centering
	\subfigure[]{\includegraphics[scale=0.26]{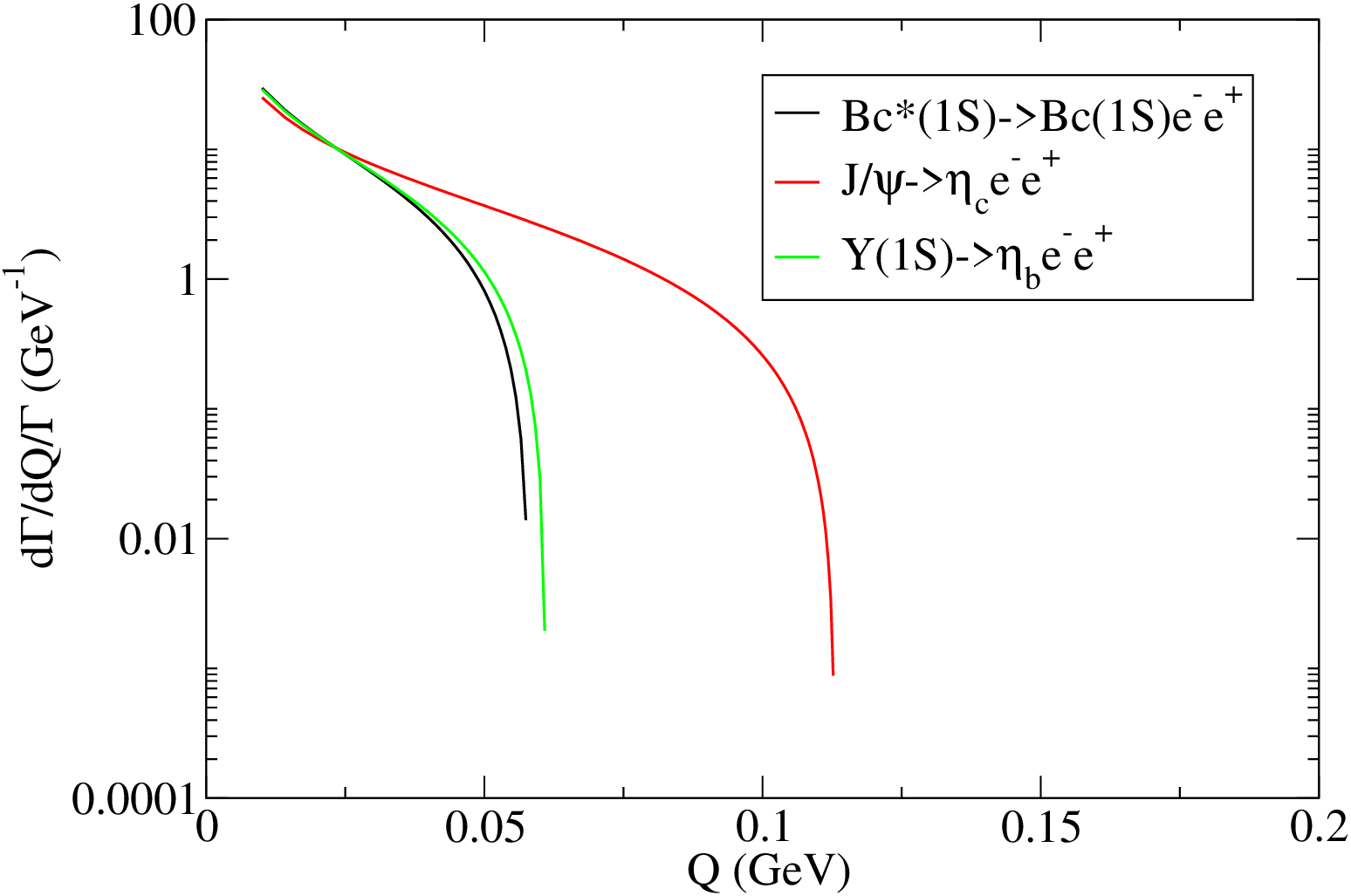}}
	\hspace{8mm}
	\subfigure[]{\includegraphics[scale=0.26]{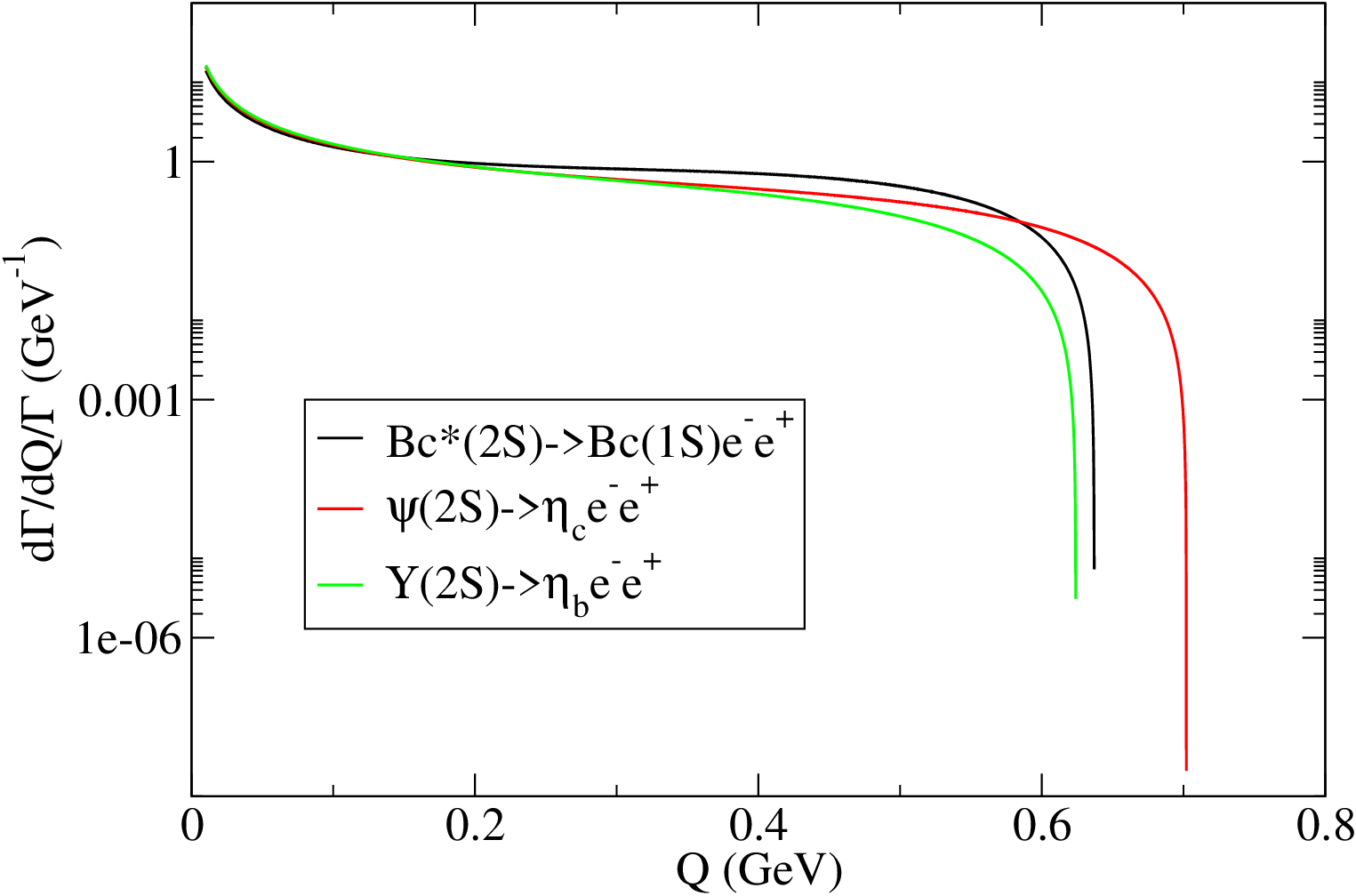}}
	\subfigure[]{\includegraphics[scale=0.26]{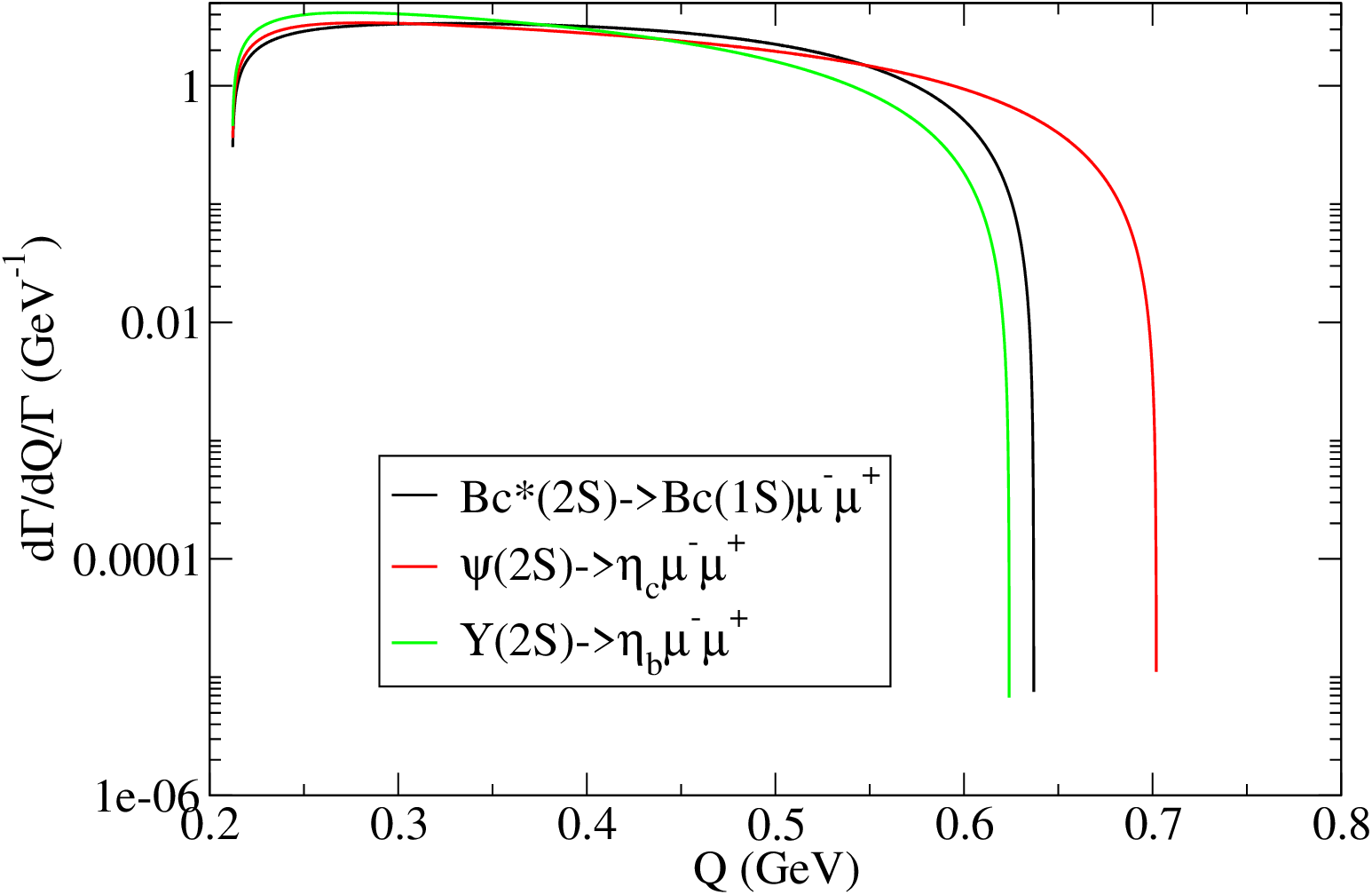}}
\caption{\label{partial width}Differential partial width (scaled by $\Gamma$) changing with $Q$.}
\end{figure}

According to Eq.~(\ref{Hmu1}) and Eq.~(\ref{Hmu2}),
the hadronic transition matrix element is related to the form factor $t$,
which is the function of the three momentum of final meson.
In Figure.~\ref{formfactor}, we have presented the numerical value of the form factor $t$ as a function of the three momentum of final meson.
The form factors are almost constant for the three decay channels of $1S$-state vector heavy mesons.
The form factors $t(J/\psi \to \eta_c)\approx 0.74$, $t(B_c^*(1S) \to B_c(1S))\approx 0.28$ and $t(\Upsilon(1S) \to \eta_b)\approx 0.14$, respectively.
But the form factors decrease as $P_f$ increase for the three decay channels of $2S$-state vector heavy mesons.
And the form factors of the $2S$-state vector heavy mesons
are much smaller than the corresponding results of the $1S$-state vector heavy mesons at zero momentum,
which reflects the suppression induced by the node of the 2S radial wave function.
Taking the form factors to Eq.~(\ref{M}) and Eq.~(\ref{Gamma1}),
we get the differential partial widths $\frac{{\rm d}\Gamma}{\Gamma{\rm d}Q}$ for the corresponding electromagnetic Dalitz decay of vector double heavy mesons.
The results are plotted in Figure.~\ref{partial width}.
The final states of $B_c^*$, $J/\psi$ and $\Upsilon(1S)$ only contain the electron-positron pairs, because of the small phase space.
And the final states of $B_c^*(2S)$, $\psi(2S)$ and $\Upsilon(2S)$ contain the electron-positron pairs and muon pairs for the large phase space.

After finishing the phase space integral, we get the partial widths of electromagnetic Dalitz decay $V \to P l^-l^+$ ($l = e,\mu$) for the
charmonium, bottomonium, and $B_c$ systems,
and the results are presented in Table~\ref{partial decay widths}. For comparison, we also present the relevant results reported in Ref.~\cite{yangmaozhi}.
The decay widths of $B_c^\ast(1S)$ and $B_c^\ast(2S)$ are all of order $10^{-10}$ GeV.
The decay widths of charmonium are larger than the corresponding results of bottomonium,
for example, $\frac{\Gamma(J/\psi\to\eta_c e^-e^+)}{\Gamma(\Upsilon(1S)\to\eta_b e^-e^+)}\approx 220$
and
$\frac{\Gamma(\psi(2S)\to\eta_c e^-e^+)}{\Gamma(\Upsilon(2S)\to\eta_b e^-e^+)}\approx 240$,
because of the charge of quarks ($e_c^2/e_b^2 = 4$) and the stronger suppression of the
wave-function overlap and phase space in the heavier bottomonium system.
The decay widths of $J/\psi\to\eta_ce^-e^+$ and $\psi(2S)\to \eta_ce^-e^+$ are consistent with the results of Ref.~\cite{yangmaozhi}.
The decay width of $\psi(2S)\to \eta_c\mu^-\mu^+$ is close to the result of Ref.~\cite{yangmaozhi}.
The decay widths of $\Upsilon(2S)\to\eta_be^-e^+$ and $\Upsilon(2S)\to\eta_b\mu^-\mu^+$ are smaller than the result of Ref.~\cite{yangmaozhi},
while remaining at the same order of magnitude, and we also give the result of $\Upsilon(1S)\to\eta_b e^-e^+$.

\begin{table}[htb]
	\centering
	\caption{The partial decay widths (in units of GeV) of di-leptonic decays $V\to Pl^-l^+$.}
	\label{partial decay widths}
    \setlength{\tabcolsep}{0.7cm}
	\begin{tabular}{ccc}
		\hline\hline
		Channel&Width(This work)  &Ref.~\cite{yangmaozhi} \\ \hline
         $B_c^\ast(1S)\to B_c(1S)e^-e^+$  &$1.86\times10^{-10}$  &-- \\
		 $J/\psi\to\eta_ce^-e^+$  & $1.06\times10^{-8}$ & $(9.6\pm2.3)\times10^{-9}$  \\
         $\Upsilon(1S)\to\eta_b e^-e^+$  & $4.76\times10^{-11}$ & --\\
         $B_c^\ast(2S)\to B_c(1S)e^-e^+$  &$6.63\times10^{-10}$ &--\\
		$\psi(2S)\to \eta_ce^-e^+$  &$7.50\times10^{-9}$ & $(8.9\pm1.3)\times10^{-9}$ \\
        $\Upsilon(2S)\to\eta_be^-e^+$  &$3.10\times10^{-11}$  & $(10.8\pm4.2)\times10^{-11}$\\
        $B_c^\ast(2S)\to B_c(1S)\mu^-\mu^+$  &$1.43\times10^{-10}$ & --\\
		$\psi(2S)\to \eta_c\mu^-\mu^+$  &$1.18\times10^{-9}$ &$(8.2\pm1.2)\times10^{-10}$ \\
		$\Upsilon(2S)\to\eta_b\mu^-\mu^+$  &$3.90\times10^{-12}$ &$(8.2\pm3.2)\times10^{-12}$\\	
		\hline\hline
	\end{tabular}
\end{table}

To facilitate direct comparison with experimental measurements,
the corresponding branching ratios are summarized in Table~\ref{branching ratios}.
We find that the
$J/\psi$ and $\psi(2S)$ channels, with branching fractions of order
$10^{-5}$--$10^{-4}$, are within the reach of the large data samples at BESIII.
The BESIII Collaboration reported the latest value for $J/\psi\to\eta_ce^-e^+$,
and our result
$\mathcal{B}(J/\psi\to\eta_c e^-e^+) = 1.15\times10^{-4}$ agrees with the measured value
$(0.92\pm0.03_{\rm stat}\pm0.10_{\rm syst}\pm0.09_{\rm input})\times10^{-4}$~\cite{BESIII:2026},
our prediction for $\mathcal{B}(\psi(2S)\to\eta_c e^-e^+)=2.62\times10^{-5}$,
which is close to the experimental value
$(3.77\pm0.40\pm0.18)\times10^{-5}$~\cite{BESIII:2022jde}.
The bottomonium channels ($\sim 10^{-7}$) may be accessible at Belle~II.
The $B_c$ channels remain challenging and will require the statistics of LHCb
or future facilities.
\begin{table}[htb]
\centering
\caption{\label{branching ratios}The branching ratios of di-leptonic decays $V\to Pl^-l^+$.}
\setlength{\tabcolsep}{0.3cm}
\begin{tabular}{ccc}
\hline\hline
Channel  & Branching ratio & Experimental value\\ \hline
        $B_c^\ast(1S)\to B_c(1S)e^-e^+$ &--&-- \\
		$J/\psi\to\eta_ce^-e^+$ & $1.15\times10^{-4}$&$(0.92\pm 0.03_{stat}\pm 0.10_{syst}\pm0.09_{input})\times 10^{-4}$\cite{BESIII:2026}  \\	
        $\Upsilon(1S)\to\eta_b e^-e^+$ & $8.82\times10^{-7}$&--\\
        $B_c^\ast(2S)\to B_c(1S)e^-e^+$ &--&--\\
		$\psi(2S)\to \eta_ce^-e^+$ & $2.62\times10^{-5}$&$(3.77\pm0.40\pm0.18)\times 10^{-5}$ \cite{BESIII:2022jde} \\
        $\Upsilon(2S)\to\eta_be^-e^+$& $9.68\times10^{-7}$&--\\
        $B_c^\ast(2S)\to B_c(1S)\mu^-\mu^+$ &--&-- \\
		$\psi(2S)\to \eta_c\mu^-\mu^+$  & $4.11\times10^{-6}$&-- \\
		$\Upsilon(2S)\to\eta_b\mu^-\mu^+$ & $1.22\times10^{-7}$&--\\
		\hline\hline
\end{tabular}
\end{table}

\section{Summary}

In this paper, we have studied the electromagnetic Dalitz decay processes
$V \to P l^-l^+$ ($l = e,\mu$) of the vector double-heavy mesons($J/\psi$, $\psi(2S)$, $\Upsilon(1S)$, $\Upsilon(2S)$, $B_c^*(1S)$ and
$B_c^*(2S)$) to pseudoscalar mesons,
in the framework of the improved Bethe--Salpeter method, which
the bound states are described by relativistic wave functions with definite
quantum numbers.

We study the transition form factors as functions of the
final-state meson momentum. For the $1S$-state vector heavy mesons transitions, the form factors
are almost constant, whereas
for the $2S$-state vector heavy mesons transitions they decrease visibly with the momentum.
The form factors of $2S$-state vector heavy mesons are much smaller than those of $1S$-state vector heavy mesons at zero
momentum, which reflects the suppression induced by the node of the $2S$
radial wave function. The charmonium form factors and widths are the largest, the $B_c$'s
results lie in between, and the bottomonium ones are the smallest, which can
be traced to the quark-charge factor and the wave-function overlap.

We give the corresponding electromagnetic Dalitz partial decay widths and branching ratios
for the vector double-heavy mesons($J/\psi$, $\psi(2S)$, $\Upsilon(1S)$, $\Upsilon(2S)$, $B_c^*(1S)$ and $B_c^*(2S)$).
For the $B_c$ channels, which have not been studied before, we predict
partial widths of order $10^{-10}$~GeV.
For charmonium channels, the branching fractions $\mathcal{B}(\psi(2S)\to\eta_c e^-e^+) = 2.62\times10^{-5}$
and $\mathcal{B}(J/\psi\to\eta_c e^-e^+) = 1.15\times10^{-4}$ agree with the
latest BESIII measurements~\cite{BESIII:2022jde,BESIII:2026}, respectively.
We also give the results of bottomonium.
We hope these results will be tested
by future experiments at BESIII, Belle~II, and LHCb.

\acknowledgments
This work was supported in part by the National Natural Science
Foundation of China (NSFC) under Grant No.~12365013, 12375085 and 12575097, the Natural Science Foundation of Ningxia
(2023AAC03276).

\appendix
\section{Instantaneous Bethe-Salpeter Equation}

In this section, we briefly review the Bethe-Salpeter equation and
its instantaneous one, the Salpeter equation.

The BS equation is read as~\cite{BS}:
\begin{equation}
(\not\!{p_{1}}-m_{1})\chi(q)(\not\!{p_{2}}+m_{2})=
i\int\frac{d^{4}k}{(2\pi)^{4}}V(P,k,q)\chi(k)\;, \label{eq1}
\end{equation}
where $\chi(q)$ is the BS wave function, $V(P,k,q)$ is the
interaction kernel between the quark and antiquark, and $p_{1},
p_{2}$ are the momentum of the quark 1 and anti-quark 2.

We divide the relative momentum $q$ into two parts,
$q_{\parallel}$ and $q_{\perp}$,
$$q^{\mu}=q^{\mu}_{\parallel}+q^{\mu}_{\perp}\;,$$
$$q^{\mu}_{\parallel}\equiv (P\cdot q/M^{2})P^{\mu}\;,\;\;\;
q^{\mu}_{\perp}\equiv q^{\mu}-q^{\mu}_{\parallel}\;.$$

In instantaneous approach, the kernel $V(P,k,q)$ takes the simple
form~\cite{Salp}:
$$V(P,k,q) \Rightarrow V(|\vec k-\vec q|)\;.$$

Let us introduce the notations $\varphi_{p}(q^{\mu}_{\perp})$ and
$\eta(q^{\mu}_{\perp})$ for three dimensional wave function as
follows:
\begin{equation}
\begin{aligned}
\varphi_{p}(q^{\mu}_{\perp})&\equiv i\int
\frac{dq_{p}}{2\pi}\chi(q^{\mu}_{\parallel},q^{\mu}_{\perp}),\\
\eta(q^{\mu}_{\perp})&\equiv\int\frac{dk_{\perp}}{(2\pi)^{3}}
V(k_{\perp},q_{\perp})\varphi_{p}(k^{\mu}_{\perp})\;. \label{eq5}
\end{aligned}
\end{equation}
Then the BS equation can be rewritten as:
\begin{equation}
\chi(q_{\parallel},q_{\perp})=S_{1}(p_{1})\eta(q_{\perp})S_{2}(p_{2})\;.
\label{eq6}
\end{equation}
The propagators of the two constituents can be decomposed as:
\begin{equation}
S_{i}(p_{i})=\frac{\Lambda^{+}_{ip}(q_{\perp})}{J(i)q_{p}
+\alpha_{i}M-\omega_{i}+i\epsilon}+
\frac{\Lambda^{-}_{ip}(q_{\perp})}{J(i)q_{p}+\alpha_{i}M+\omega_{i}-i\epsilon}\;,
\label{eq7}
\end{equation}
with
\begin{equation}
\omega_{i}=\sqrt{m_{i}^{2}+q^{2}_{_T}}\;,\;\;\;
\Lambda^{\pm}_{ip}(q_{\perp})= \frac{1}{2\omega_{ip}}\left[
\frac{\not\!{P}}{M}\omega_{i}\pm
J(i)(m_{i}+{\not\!q}_{\perp})\right]\;, \label{eq8}
\end{equation}
where $i=1, 2$ for quark and anti-quark, respectively,
 and
$J(i)=(-1)^{i+1}$.

Introducing the notations $\varphi^{\pm\pm}_{p}(q_{\perp})$ as:
\begin{equation}
\varphi^{\pm\pm}_{p}(q_{\perp})\equiv
\Lambda^{\pm}_{1p}(q_{\perp})
\frac{\not\!{P}}{M}\varphi_{p}(q_{\perp}) \frac{\not\!{P}}{M}
\Lambda^{{\pm}}_{2p}(q_{\perp})\;. \label{eq10}
\end{equation}
With contour integration over $q_{p}$ on both sides of
Eq.~(\ref{eq6}), we obtain:
$$
\varphi_{p}(q_{\perp})=\frac{
\Lambda^{+}_{1p}(q_{\perp})\eta_{p}(q_{\perp})\Lambda^{+}_{2p}(q_{\perp})}
{(M-\omega_{1}-\omega_{2})}- \frac{
\Lambda^{-}_{1p}(q_{\perp})\eta_{p}(q_{\perp})\Lambda^{-}_{2p}(q_{\perp})}
{(M+\omega_{1}+\omega_{2})}\;,
$$
and the full Salpeter equation:
\begin{equation}
\begin{aligned}
&(M-\omega_{1}-\omega_{2})\varphi^{++}_{p}(q_{\perp})=
\Lambda^{+}_{1p}(q_{\perp})\eta_{p}(q_{\perp})\Lambda^{+}_{2p}(q_{\perp}),\\
&(M+\omega_{1}+\omega_{2})\varphi^{--}_{p}(q_{\perp})=-
\Lambda^{-}_{1p}(q_{\perp})\eta_{p}(q_{\perp})\Lambda^{-}_{2p}(q_{\perp}),\\
&~~~~~~~~~~~~~~~~\varphi^{+-}_{p}(q_{\perp})=\varphi^{-+}_{p}(q_{\perp})=0.
\label{eq11}
\end{aligned}
\end{equation}

For the different $J^{PC}$ (or $J^{P}$) states, we give the general form of
wave functions. Reducing the wave functions by the last
equation of Eq.~(\ref{eq11}), then solving the first and second equations in Eq.~(\ref{eq11}) to
get the wave functions and mass spectrum. We have discussed the
solution of the Salpeter equation in detail in Ref.~\cite{mass1,w1}.

The normalization condition for BS wave function is:
\begin{equation}
\int\frac{q_{_T}^2dq_{_T}}{2{\pi}^2}Tr\left[\overline\varphi^{++}
\frac{{/}\!\!\!
{P}}{M}\varphi^{++}\frac{{/}\!\!\!{P}}{M}-\overline\varphi^{--}
\frac{{/}\!\!\! {P}}{M}\varphi^{--}\frac{{/}\!\!\!
{P}}{M}\right]=2P_{0}\;. \label{eq12}
\end{equation}

 In our model, the instantaneous interaction kernel $V$ is Cornell
potential, which is the sum of a linear scalar interaction and a vector interaction:
\begin{equation}\label{vrww}
V(r)=V_s(r)+V_0+\gamma_{_0}\otimes\gamma^0 V_v(r)= \lambda
r+V_0-\gamma_{_0}\otimes\gamma^0\frac{4}{3}\frac{\alpha_s}{r}~,
\end{equation}
 where $\lambda$ is the string constant and $\alpha_s(\vec
q)$ is the running coupling constant. In order to fit the data of
heavy quarkonia, a constant $V_0$ is often added to confine
potential. One can see that $V_v(r)$ diverges at $r=0$, we introduce a factor $e^{-\alpha r}$ to avoid
the divergence:
\begin{equation}
V_s(r)=\frac{\lambda}{\alpha}(1-e^{-\alpha r})~,
~~V_v(r)=-\frac{4}{3}\frac{\alpha_s}{r}e^{-\alpha r}~.
\end{equation}\label{vsvv}

 It is easy to
know that when $\alpha r\ll1$, the potential becomes to Eq.~(\ref{vrww}). In the momentum space and the C.M.S of the bound state,
the potential reads :
\begin{equation}
\begin{aligned}
&~~~~~~~~~~~~~~~~~~~~~~~~~~V(\vec q)=V_s(\vec q)
+\gamma_{_0}\otimes\gamma^0 V_v(\vec q),\\
&V_s(\vec q)=-(\frac{\lambda}{\alpha}+V_0) \delta^3(\vec
q)+\frac{\lambda}{\pi^2} \frac{1}{{(\vec q}^2+{\alpha}^2)^2},
~~V_v(\vec q)=-\frac{2}{3{\pi}^2}\frac{\alpha_s( \vec q)}{{(\vec
q}^2+{\alpha}^2)},\label{eq16}
\end{aligned}
\end{equation}
where the running coupling constant $\alpha_s(\vec q)$ is :
$$\alpha_s(\vec q)=\frac{12\pi}{33-2N_f}\frac{1}
{\log (a+\frac{{\vec q}^2}{\Lambda^{2}_{QCD}})}~.$$

We introduce a small
parameter $a$ to
avoid the divergence in the denominator. The constants $\lambda$, $\alpha$, $V_0$ and
$\Lambda_{QCD}$ are the parameters that characterize the potential. $N_f=3$ for $\bar bq$ (and $\bar cq$) system.

\end{document}